\documentclass[prd,a4paper,showpacs,preprint,byrevtex]{revtex4}
\usepackage{graphicx}
\usepackage{dcolumn}
\usepackage{array}

\usepackage{amssymb}
\usepackage{amsmath}
\usepackage{amsfonts}
\usepackage{bm}

\begin{document}

\title{Modified Newton's law, brane worlds, and the gravitational quantum well}

\author{Fabien Buisseret}
\thanks{FNRS Research Fellow}
\email[E-mail: ]{fabien.buisseret@umh.ac.be}
\affiliation{Groupe de Physique Nucl\'{e}aire Th\'{e}orique,
Universit\'{e} de Mons-Hainaut,
Acad\'{e}mie universitaire Wallonie-Bruxelles,
Place du Parc 20, BE-7000 Mons, Belgium}
\author{Bernard Silvestre-Brac}
\affiliation{Laboratoire de Physique Subatomique et de Cosmologie, Avenue des Martyrs 53, 38026 Grenoble-Cedex, France}
\email[E-mail: ]{silvestre@lpsc.in2p3.fr}
\author{Vincent Mathieu }
\thanks{IISN Scientific Research Worker}
\email[E-mail: ]{vincent.mathieu@umh.ac.be}
\affiliation{Groupe de Physique Nucl\'{e}aire Th\'{e}orique,
Universit\'{e} de Mons-Hainaut,
Acad\'{e}mie universitaire Wallonie-Bruxelles,
Place du Parc 20, BE-7000 Mons, Belgium}

\date{\today}

\begin{abstract}
Most of the theories involving extra dimensions assume that only the gravitational interaction can propagate in them. In such approaches, called brane world models, the effective, $4$-dimensional, Newton's law is modified at short as well as at large distances. Usually, the deformation of Newton's law at large distances is parametrized by a Yukawa potential, which arises mainly from theories with compactified extra dimensions. In many other models however, the extra dimensions are infinite. These approaches lead to a large distance power-law deformation of the gravitational
newtonian potential $V_N(r)$, namely $V(r)=(1+k_b/r^b)V_N(r)$, which is less studied in the literature. We investigate here the dynamics of a particle in a gravitational quantum well with such a power-law deformation. The effects of the deformation on the energy spectrum are discussed. We also compare our modified spectrum to the results obtained with the GRANIT experiment, where the effects of the Earth's gravitational field on quantum states of ultra cold neutrons moving above a mirror are studied. This comparison leads to upper bounds on $b$ and $k_b$. 
\end{abstract}



\maketitle

\section{Introduction}\label{intro}
Since the pioneering works of Kaluza and Klein \cite{KK}, which aimed to unify gravitation and electromagnetism in the framework of a $5$-dimensional model, theories involving extra dimensions have been considerably studied in theoretical physics. In particular, brane world models, introduced in the eighties \cite{brane}, are a very active field of research. Although there is a great variety of them~\cite{brane2,Bronnikov:2006jy}, they all rely on the common idea that our $4$-dimensional universe is a subspace (a brane) of a larger, higher dimensional, universe (the bulk). The extra dimensions can either be compactified, or infinite, and the number of extra-dimensions, as well as the fields living in the bulk, vary from one model to another. One of their most interesting feature is that brane world models are good candidates to solve the so-called hierarchy problem, that is the unexplained weakness of the gravitational interaction with respect to other forces. 

A first suggestion made by Arkani-Hamed \emph{et al} \cite{ADD} in order to solve the hierarchy problem with brane world models is that our universe could be a $(4+n)$-dimensional manifold, with $n$ compactified extra dimensions \cite{ADD,RS1}. Their basic assumption is that, while all the standard model fields - gauge and matter - are confined in the usual $4$ dimensions, only the graviton can propagate freely in the $(4+n)$ dimensions, eventually leading to a very weak ``effective" gravitational interaction in the $4$-dimensional universe. Thus, if the compactification length scale of the extra dimensions is, say, $r_c$, gravity will only become relevant with respect to the other forces at scales lower than $r_c$. Brane world models with compact extra dimensions, also known as Large Extra Dimensions (LED) models, and their possibly observable consequences, have been considerably studied in the literature~\cite{led}. As an example, let us consider that the universe possesses $n$ LED, compactified on a torus whose $n$ radii are equal to $r_c$. Then, for distances $r\gg r_c$, the newtonian potential 
\begin{equation}\label{potnew}
V_N(r)=-\frac{GmM}{r}	
\end{equation}
between two pointlike particles is modified to the following form~\cite{Keh} 
\begin{equation}\label{Yuk}
	V_{LED}(r)=V_N(r)\left(1+\alpha{\rm e}^{-\lambda r}\right),\quad r\gg r_c,
\end{equation}
with $\alpha=2n$, $\lambda=1/r_c$. For distances smaller than $r_c$ however, gravitation is enhanced since the newtonian potential becomes~\cite{ADD}
\begin{equation}\label{sr}
	V_{LED}(r)=V_N(r)\left(\frac{r_c}{r}\right)^n,\quad r\ll r_c.	
\end{equation}
Nowadays, precise constraints on the possible values of $\alpha$ and $\lambda$ in the general Yukawa form (\ref{Yuk}) have been obtained from various experiments \cite{Mesu}. In the particular case of the LED approach, these constraints are upper bounds on $r_c$~\cite{Hall,Hanne,Abbo}, but in general they can be applied to various theoretical models predicting a change in the gravitational law~\cite{long1,long2}.   
\par Another approach to solve the hierarchy problem has been proposed by Randall and Sundrum in Ref.~\cite{RS2}. Their proposal was also based on brane world models, but they allowed for the extra dimension to be infinite. As the LED approaches, brane world models with infinite extra dimensions lead to a modification of Newton's law. However, instead of the Yukawa form (\ref{Yuk}), these modifications at large distances are generally of the form \cite{Bronnikov:2006jy} 
\begin{equation}\label{modif}
	V(r)=V_N(r)\left(1+\frac{k_b}{r^b}\right),\quad r\gg\Lambda=\sqrt[b]{|k_b|},
\end{equation}
where $\Lambda$ can be considered as a typical length scale at which the correction due to the infinite extra dimensions becomes dominant. For $r\ll\Lambda$ however, $V(r)\propto r^{-(n+1)}$ as in Eq.~(\ref{sr}), even in this case where the $n$ extra dimensions are infinite. It appears that a power-law parametrization such as Eq.~(\ref{modif}) is poorly studied in the literature compared to the Yukawa deformation. Consequently, we think that a discussion of its physical consequences is of interest. To this purpose, we propose to investigate the effect of brane world corrections such as those of Eq.~(\ref{modif}) on a particular physical system: The gravitational quantum well (GQW), i. e., a bound state of a particle in the gravitational potential.
\par A recent experiment, called GRANIT, is devoted to the study of quantum states of neutrons in the Earth's gravitational field. Roughly speaking, in this experiment, ultra cold neutrons are freely moving (bouncing) in the gravitational field above a mirror. This particular setup gives rise to a GQW. As a consequence, the energy spectrum of the neutrons in the gravitational field's direction is quantized, and the probability of observing a particle at a given height will be maximum at the classical turning points $h_n=E_n/mg$, for each energy $E_n$. This is precisely what is observed. More details can be found in Refs~\cite{nesv02,nesv05}. This experiment gives an opportunity to make a confrontation between observation and various theoretical models concerning quantum effects in gravity: Noncommutative geometry \cite{berto}, existence of an intrinsic minimal length \cite{brau}, spin-gravity effects \cite{Hehl,FabB},\dots Let us note that a modification of Newton's law of the Yukawa form (\ref{Yuk}) has already been studied in Ref.~\cite{GRA} by using the GRANIT data. Here, we will follow a similar way to derive constraints on the corrections~(\ref{modif}).
\par Our paper is organized as follows. In Sec.~\ref{gqw}, we present the basic properties of the gravitational quantum well. Then, we compute the effective potential acting on the bouncing particle in Sec.~\ref{ed_newton}, that is the usual newtonian potential plus a term coming from brane world corrections. In Sec.~\ref{constr} we discuss the effects of such corrections on the GQW spectrum and we compute constraints on the correction parameters by comparison with the GRANIT experiment. Finally, we draw some conclusions in Sec.~\ref{conclu}.

\section{The gravitational quantum well}\label{gqw}
Let us consider the case of a nonrelativistic particle of mass $m$, subject to a gravitational field given by $\vec{g}=-g\, \vec{e}_z$, with $g$ a constant. In order to form a GQW, a mirror is placed at $z=0$ and acts as an hardcore interaction. This corresponds to the experimental setup described in Refs~\cite{nesv02,nesv05}. The corresponding potential is $V_0(\vec{x}\,)=V_0(z)$ with
\begin{eqnarray}
\label{eq4a}
V_0(z)&=& +\infty \quad {\rm for} \quad z\le 0 \nonumber \\
    &=& mg z \quad {\rm for} \quad z>0.
\end{eqnarray}
An infinite potential in $z=0$ is a very good description of the mirror, at least for the lowest eigenstates. The simplest model describing the vertical motion of the bouncing particle at the quantum level is then a Schr\"odinger equation with a linear potential \begin{equation}\label{simple}
H_0\psi(z)=E_0\psi(z),	
\end{equation}
with
\begin{equation}
H_0=\frac{p^2}{2m}+mgz.
\end{equation}
The mirror, located in $z=0$, imposes the boundary condition $\psi(0)=0$ corresponding to the infinite potential barrier. This condition causes the energy to be quantized. The solution of such a problem is well-known \cite[Problem 40]{Flu}, and reads
\begin{equation}
	E_{0n}=-\left(\frac{mg}{\theta}\right) \alpha_n,
\end{equation}
\begin{equation}\label{psi0}
	\psi_n(z)=N_n {\rm Ai} \left[\theta\, z+\alpha_n\right],
\end{equation}
where
\begin{equation}\label{thetadef}
	\theta=\left(\frac{2m^2 g}{\hbar^2}\right)^{1/3}.
\end{equation}
$\alpha_n$ is the $n^{\rm{th}}$ zero of the regular Airy function Ai$(z)$ (these zeros can be found for example in Ref.~\cite{abra}), and $N_n=\theta^{1/2}/|{\rm Ai}'(\alpha_n)|$ is the normalization factor. In analogy with the classical situation, a particle of energy $E_n$ should bounce at a height given by 
\begin{equation}\label{eq3}
	h_n=\frac{E_n}{mg}=-\frac{\alpha_n}{\theta}.
\end{equation}
In the GQW, the critical heights are thus quantized. This has been checked by the GRANIT experiment \cite{nes}, in which ultra cold neutrons are moving above a mirror in the Earth's gravitational field. The experimentally measured critical heights for the first two states are 
\begin{eqnarray}
	h^{\rm{exp}}_1&=&12.2 \rm{\mu m}\pm1.8_{\rm{syst}}\pm0.7_{\rm{stat}},\\
	h^{\rm{exp}}_2&=&21.6 \rm{\mu m}\pm2.2_{\rm{syst}}\pm0.7_{\rm{stat}},
\end{eqnarray}
while formula~(\ref{eq3}) with $m = 939$ MeV$/$c$^2$ and $g=9.81$ m$/$s$^2$ gives
\begin{eqnarray}
	h^{\rm{th}}_1&=&13.7 \rm{\mu m},\\
	h^{\rm{th}}_2&=&24.0 \rm{\mu m}.
\end{eqnarray}
\par The theoretical values are located within the error bars, showing the validity of our simple Hamiltonian approach. Actually, as the neutrons are removed from any magnetic field in the GRANIT experiment, the main corrections that we could miss with the Schr\"odinger equation~(\ref{simple}) are spin-gravity corrections, due to the fermionic nature of the neutron. Such corrections are of the form $\vec S(\vec g\times\vec p)/2mc^2$~\cite{Hehl}, but they are too small to be observable. As a consequence of the good agreement between theory and experiment, the energy shifts due to eventual new physical mechanisms must be bounded. A similar argument has already been used to derive constraints on theories with a noncommutative geometry~\cite{berto}, or with a minimal length uncertainty~\cite{brau}. As a possible way to derive constraints on new contributions, namely power-law modifications of the gravitational potential, we propose here to study the variation of $\delta=h_2-h_1$ when these contributions are added. Theoretically, we obtain $\delta^{{\rm th}}=10.3\ \mu$m, while the experimental value is $\delta^{{\rm exp}}=9.4\, \mu$m$\, \pm 5.4\, \mu$m. If we assume that the potential in Eq.~(\ref{simple}) is no longer $mgz$ but $mgz+U(z)$, with $U(z)$ a perturbation term encoding eventual new physics, then $\delta$ will be changed into $\delta+\Omega$ , with 
\begin{equation}\label{omegadef1}
	\Omega=\frac{1}{mg}\left[\left\langle \psi_2\right|U\left|\psi_2\right\rangle-\left\langle \psi_1\right|U\left|\psi_1\right\rangle\right],
\end{equation}
the eigenstates being given by formula~(\ref{psi0}). The comparison with experimental data requires that $\Omega$ satisfies
\begin{equation}\label{omegadef2}
|\Omega|\leq4.5\ \mu {\rm m}.
\end{equation}
From this relation, inequalities ruling the parameters of the perturbation term can be derived, as we will show in Sec.~\ref{constr}. Let us point out that working with $\Omega$ is useful because it avoids to deal with eventual arbitrary constants which could appear in $U(z)$. 

\section{Power-law corrections to Newton's law}\label{ed_newton}
As we already mentioned, all the theoretical approaches involving extra dimensions predict a modification of Newton's law. At large distances, the LED framework predicts that these deviations should be parametrized by the Yukawa potential
\begin{equation}
V(r)=-\frac{ Gm  M}{r}\left(1+\alpha\, {\rm e}^{-r/r_c} \right),\quad r\gg r_c.
\end{equation}
As the previous deformation has already been studied in Ref.~\cite{GRA}, we will rather consider in this work the following form for the modified Newton's law, 
\begin{equation}\label{potpuiss}
	V(r)=-\frac{Gm M }{r}\left(1+\frac{k_b}{r^b}\right),\quad r\gg\sqrt[b]{|k_b|},
\end{equation}
appearing in the long range corrections due to brane world models with infinite extra dimensions.
If $V_0(r)=-GMm/r$ is the usual newtonian potential, we can rewrite Eq.~(\ref{modif}) as $V(r)=V_0(r)+V_b(r)$. In this picture, $\Lambda=\sqrt[b]{|k_b|}$ is a typical length scale at which the correction becomes dominant. Eventually, we will apply our results to the GRANIT experiment. As it can be computed from Sec.~\ref{gqw}, a typical size of this experiment is around $1/\theta\approx 6\, \mu$m. As, in every case, the modifications of Newton's law due to extra dimensions are expected to become dominant at length scales well below the micrometer ($\Lambda\ll 1/\theta$), we assume that our analysis is the domain of validity of formula~(\ref{potpuiss}).

\subsection{Earth-particle interactions}
\par Potential (\ref{potpuiss}) describes an interaction between two pointlike particles. However, in our case, we are dealing with a pointlike particle bouncing on a plane mirror, at the surface of the Earth. Consequently, we have to derive the effective potentials between our particle and the Earth on one side, and between the particle and the mirror on the other side. Although the mirror's mass is negligible with respect to the Earth's one, its influence should also be taken into account because it is much closer to the particle. Firstly, we begin with the Earth-particle interactions. Our Planet can be seen in first approximation as a spherical body whose density $\rho_E$ is constant, and whose radius is $R=6378$ km. If the potential between the particle and an infinitesimal Earth's element $dM=\rho_E\, d^3 \vec{x}'$ is given by
\begin{equation}
	dV_b(\vec{x})=-\frac{m G  \rho_E k_b}{|\vec{x}-\vec{x}^{\, '}|^{b+1}}d^3\vec{x}^{\, '},
\end{equation}
then, the total potential acting on the particle is 
\begin{equation}
	V_b(\vec{x})=-m G  \rho_E k_b\int_{E}\frac{d^3\vec{x}^{\, '}}{|\vec{x}-\vec{x}^{\, '}|^{b+1}},
\end{equation}
with $E$ the volume the Earth.
\begin{figure}[t]
\begin{center}
\includegraphics*[height=4cm]{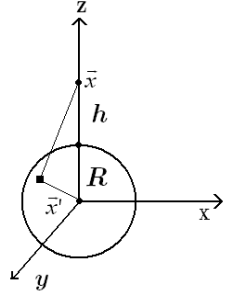}
\end{center}
\caption{Schematic view of the particle, located at a height $h$ above the Earth's surface.}
\label{fig:n1}
\end{figure}
It can be deduced from Fig.~(\ref{fig:n1}) that, in spherical coordinates, 
\begin{equation}\label{pot1}
V_b(h)=-2\pi m G \rho_E k_b\int^R_0 r^2 dr	\int^1_{-1} \frac{du}{\left[r^2-2r(h+R)u+(h+R)^2\right]^{(b+1)/2}}\ ,
\end{equation}
where $h$ is the altitude above Earth's surface. For all real values of $b\neq1,2,3$, we obtain 
\begin{equation}
V_{b}(h)=-\frac{3  m g k_{b}}{2(b-1)(b-2)(b-3)R(h+R)}\left[\frac{(b-3)R-h}{h^{b-2}}+\frac{(b-1)R+h}{(h+2R)^{b-2}}\right],
\end{equation}
where the relations $M=(4/3)\pi R^3\rho_E$ and $g=GM/R^2$ were used. As a test, we can check that we recover the correct Newton's law for $k_0=1$,
\begin{equation}
V_0(h)=-\frac{GmM}{h+R}.	
\end{equation}
For $b=1,2,3$, the calculation needs a special treatment but the results are still analytical
\begin{eqnarray}
	V_1(h)&=&-\frac{3mgk_1}{4R}\left[2R-\frac{h(h+2R)}{h+R}\ln\left(\frac{h+2R}{h}\right)\right],\\
	V_2(h)&=&-\frac{3mgk_2}{2R}\left[\ln\left(\frac{h+2R}{h}\right)-\frac{2R}{h+R}\right],\\
	V_3(h)&=&-\frac{3mgk_3}{4R}\left[\frac{2R}{h(h+2R)}-\frac{1}{h+R}\ln\left(\frac{h+2R}{h}\right)\right].
\end{eqnarray}
Let us note that almost all these expressions are singular in $h=0$. 
\par In our GQW, we can assume that $h/R\ll1$. By dropping the constants and keeping only the dominant terms in $h/R$, we obtain

 \begin{eqnarray}\label{pot_E}
	V^E_0(h)&\approx&mgk_0h,\\
	V^E_1(h)&\approx&-\frac{3mgk_1}{2R}\, h\ln\left(\frac{h}{2R}\right),\\
	V^E_2(h)&\approx&\frac{3mgk_2}{2R}\ln \left(\theta h\right),\\
	V^E_{b\geq3}(h)&\approx&-\frac{3  mg k_b}{2(b-1)(b-2)R}\frac{1}{h^{b-2}}.
\end{eqnarray}

When it was possible, we replaced $\ln (h/2R)$ by $\ln(\theta h)$ with $\theta$ defined by (\ref{thetadef}) because $1/\theta$ is the characteristic length scale of a GQW. The difference in both expressions is just an irrelevant constant.

\subsection{Mirror-particle interactions}

The mirror can be seen as a parallelepiped whose density is $\rho_M$, that we can parametrize as follows: $-\infty<x,y<\infty$ and $-L<z<0$. At the scale of the particle, it is indeed justified to approximate the mirror by an infinite plane. In this case, it is easier to consider the mirror as a disc with an infinite radius. The equivalent of Eq.~(\ref{pot1}) is then
\begin{equation}
V_b(h)=-2\pi m G  \rho_M k_b\int^0_{-L}dz	\int^{R^*}_0 \frac{r\:dr}{\left[r^2+(h-z)^2\right]^{(b+1)/2}},
\end{equation}
with $R^*$ very large but finite in order to avoid divergent integrals (these divergences will disappear later). We find 
\begin{equation}
	V_1(h)=\frac{3mgk_1}{2R}\rho\left[L \ln(h+L)+h \ln\left(\frac{h+L}{h}\right)\right],
\end{equation}
\begin{equation}
	V_2(h)=-\frac{3mgk_2}{2R}\rho\ln\left(\frac{h+L}{h}\right),
\end{equation}
\begin{equation}
	V_{b>2}(h)=-\frac{3mgk_b}{2R}\rho \frac{1}{(b-1)(b-2)}\left[\frac{1}{h^{b-2}}-\frac{1}{(h+L)^{b-2}}\right],
\end{equation}
with $\rho=\rho_M/\rho_E$. The constant terms involving $R^*$ have been removed, since they do not have any physical meaning. By only keeping the dominant terms in $h/L$, we have
\begin{eqnarray}\label{pot_M}
	V^M_{0}(h)&\approx&\rho V^E_{0}(h),\\
	V^M_1(h)&\approx&-\frac{3mgk_1}{2R}\rho h \ln\left(\frac{h}{L}\right),\\
	V^M_{b\geq2}(h)&\approx&\rho V^E_{b\geq 2}(h).
\end{eqnarray}

Let us mention that in the GRANIT experiment, $L\approx10$ cm and $\rho\approx1$~ \cite{GRA}.
One sees that the formulae for the mirror have exactly the same analytical form as the ones relative to the Earth. This is not surprising since in both cases the size of the neutron is much, much smaller than the size of the gravitational sources which appear to the neutron locally as an infinite disc.

\section{The modified GQW spectrum}\label{constr}

\subsection{Influence of the corrections}
We discussed in Sec.~(\ref{gqw}) how a new physical mechanism, contained in a perturbation $U(z)$, could be included in the GQW. Clearly, in our case, we have 
\begin{equation}\label{udef}
	U_b(z)=V^E_b(z)+V^M_b(z).
\end{equation}
Roughly speaking, $U_{b>0}(z)\propto -mgk_b/R$. Generally in brane world models, $k_b$ is a positive number. In this case, $U_b(z)$ will bring a negative contribution to the energy, and will thus decrease the critical heights. This is coherent with the prediction of almost all modified theories of gravitation that the effective gravitational interaction is enhanced, especially at short scales. We can remark however that recently, a new model, based in particular on Gauss-Bonnet gravity, was proposed, in which the gravitational interaction was decreased at short scale~\cite{deRham:2006hs}. In this special theory, the critical heights would be higher than with the usual newtonian potential. At least theoretically, the GQW is able to distinguish between theories predicting an enhancement or a decrease of the gravitational interaction because, in a classical picture, the particle will be able to bounce higher than predicted with the usual Newton's law in the second case, and lower in first case.  
\par The trivial case where $b=0$ does not correspond to a modification of the gravitational potential, but rather to a change in the value of the gravitational constant $G$, which is replaced by $G(1+k_0)$. It can be encountered for example in theories with supersymmetric extra dimensions \cite{Callin:2005wi}. Moreover, the calculations can be done analytically in this case as an illustration of our discussion. Thanks to the properties of the Airy function, it can be checked that
\begin{equation}
	\langle z\rangle=-\frac{2\, \alpha_n}{3\, \theta},
\end{equation}
and consequently, the level $h_n$ will be shifted by an amount
\begin{equation}
	\Delta h_n=-\frac{4k_0}{3}h_n.
\end{equation}
As we argued previously, if $k_0$ is positive (negative), it will decrease (increase) the critical heights. Note that $\Delta h_n$ also increases with $n$: The corrections logically become more important for higher energies.

Finally, let us remark that, as a level $h_n$ will be shifted by $\Delta h_n=\left\langle U_b\right\rangle/mg$, we have
\begin{equation}\label{bad1}
\Delta h_n\propto k_b/R.	
\end{equation}
This ratio should be very small in any earthly GQW experiment. Consequently, such a particular setup does not appear to be very sensitive to the brane world corrections we consider here. A confirmation of this point will be given in the next section, where we compare our results to the GRANIT experiment.

\subsection{Constraints on the power-law parameters} 
Let us define $U_b=-k_bmg\tilde U_b\equiv-\Lambda^b mg\tilde U_b$. Then, rewriting Eq.~(\ref{omegadef1}), we obtain
\begin{equation}\label{omegadef3}
	\Omega_b\equiv\Lambda^b\tilde\Omega_b=-\Lambda^b\left[\left\langle \psi_2\right|\tilde U_b\left|\psi_2\right\rangle-\left\langle \psi_1\right|\tilde U_b\left|\psi_1\right\rangle\right]. 
\end{equation}
By combining the definition (\ref{omegadef3}) and the experimental inequality (\ref{omegadef2}), coming from the GRANIT results, we are led to the following constraint on the characteristic length scale of the Newton's law modification $\Lambda$,
\begin{equation}\label{co_k}
	|\Lambda|\leq \sqrt[b]{4.5/\left|\tilde\Omega_b\right|}\ \mu{\rm m}.
\end{equation}
As the spectrum of the unperturbed GQW is well known, $\tilde\Omega_b$ can be computed for every value of $b$ in the case of a neutron in the Earth's gravitational field. The upper bounds we find on $|\Lambda|$ are summed up in Tab.~\ref{tab2}. For the calculations we used a ratio between the mirror and Earth densities equal to $\rho = 1$. In the case $b \ne 1$ this means that
the mirror gives a contribution exactly equal to that of the Earth. For the case $b=1$ we chose $L=10$ cm. Clearly, these bounds are not very good, but this was not a priori known; unfortunately, it only appears once the calculations are done. Moreover, we find useful to give them since such bounds on power-law deformations are less common in literature than bounds on Yukawa deformations. Beyond problems of experimental precision, it is somewhat amazing that values of $\Lambda$ as large as $1$ meter would not affect the GQW spectrum in any observable way, since the typical length scale of this system is only $1/\theta\approx 6\, \mu$m with the parameters of the GRANIT experiment. The GQW is thus not very sensitive to the brane world corrections that we considered, since these corrections come from effective interactions with the mirror and the Earth. As shown in Eq.~(\ref{bad1}), these are very small because of the large value of $R$. If the experimental precision is increased of a factor $10^f$, the upper bounds on $\Lambda$ will roughly be improved by a factor $10^{f/b}$, thus the improvement will be particularly interesting for the smaller values of $b$.     
\begin{table}[h]
	\centering
		\begin{tabular}{c|cccc}
$b$ & $1$ & $2$ & $3$ & $4$ \\
\hline						
$|\Lambda|$(m)$<$ & $8.13\ 10^4$ & $4.34$ &$0.076$&$0.009$  \\		
		\end{tabular}
		\caption{Constraints on the power-law parameters of Eq.~(\ref{potpuiss}). The values $\rho = 1$ and $L=10$ cm were used.}
\label{tab2}
\end{table}
\par As an illustration, we can mention some brane world models with infinite extra dimensions which cause Newton's law to be modified with Eq.~(\ref{potpuiss}). We begin with the well-known Randall-Sundrum II model \cite{RS2}, which involves one uncompactified extra dimension. It is shown in Ref.~\cite{Call} that the effective newtonian potential on the brane is
\begin{equation}\label{rspot}
	V_{{\rm RS}}=\left\{
	\begin{array}{c}
	 -\frac{GmM}{r}(1+4\Lambda_R/3\pi r)\quad r\ll \Lambda_R\\
	 -\frac{GmM}{r}(1+2\Lambda_R^2/3 r^2)\quad r\gg \Lambda_R
	\end{array}
	\right. .
\end{equation}
$\Lambda_R$ is a typical length scale of the Randall-Sundrum model. Intuitively, one could think that it has to be very small, but some recent theoretical arguments suggest that it could be about $0.1$ mm ($\approx10^{-3}$ eV ), without contradiction with the current observational evidences \cite{Eve}. The order of magnitude of $\Lambda_R$ is thus still a matter of discussion. The large $r$ limit of potential~(\ref{rspot}) corresponds to $k_2=\Lambda^2=2\Lambda_R^2/3$ respectively. We find then $\Lambda_R<5.3~\rm{m}$, which is considerably worst than the upper bound of Ref.~\cite{long1}: $\Lambda_R<0.1$~mm. We can notice that it does not exclude the arguments of Ref.~\cite{Eve}. Moreover, even if $\Lambda_R$ were equal to $0.1$ mm, it would cause $\Omega$ to be around $2$ fm, which seems clearly unobservable, even with an improvement of the experimental precision in GRANIT. Let us point out that the Gauss-Bonnet brane world model of Ref.~\cite{deRham:2006hs} enters in the category $b=2$. 
\par Another model which is of interest involves a dilaton field in a $5$-dimensional bulk, added to the usual gravity~\cite{Noji}. In this approach, the long-range newtonian potential has the form 
\begin{equation}\label{dila}
	V_{D}=-\frac{GmM}{r}\left(1+\frac{\Lambda_D^3}{r^3}\right),\quad  r\gg\Lambda_D
\end{equation}
$\Lambda_D$ depending in particular on the dilaton field \cite{Noji}. 
\par Another approach was recently proposed, which consider spherically symmetric solutions of Einstein's equations
with a bulk cosmological constant and $n$ transverse dimensions \cite{bad1,bad2}. Gravity is then localised on a four dimensional topological defect. The newtonian potential has the form
\begin{equation}\label{bad}
	V_{n}=-\frac{GmM}{r}\left[1+\frac{\Lambda^{n+1}(n)}{r^{n+1}}\right],\quad r\gg\Lambda(n).
\end{equation}
Here, the characteristic length scale $\Lambda(n)$ depends on the number of extra dimensions. If $n=3$, we see that this last model predicts a modification in $r^{-4}$ of Newton's law.
\par For what concerns theories with $b\geq 5$, an important theoretical result has to be pointed out. Let us assume, as we did in Sec.~\ref{ed_newton}, that the characteristic size of the brane world corrections, denoted as $\Lambda$, is much more smaller that the gravitational quantum well length scale, i.e. $1/\theta$. Then, Eqs~(\ref{pot_E}) and (\ref{pot_M}) will still be valid at very small $h$ with respect to the size of the GRANIT experiment. But, for $b\geq 5$, $U_b(h)$ is more divergent at small $h$ than $h^{-2}$. In consequence, if $U_{b\geq 5}$ is valid at sufficiently small $h$, as we assumed here, it will forbid the existence of bound states in the GQW, because it is well-known that a potential more singular than $h^{-2}$ never admits bound states. Even if $\Lambda$ was large enough to avoid such a collapse of the bound states, it should bring a very large negative contribution to the energy levels. This is obviously in contradiction with what is experimentally observed, since the theoretical results obtained with the unperturbed gravitational potential are close to the experimental values. We are thus led to the upper bound 
\begin{equation}\label{up}
	b < 5.
\end{equation}
For example, formula (\ref{bad}) predicts that this model should be discarded for $n>3$. 

\par Finally, let us mention that even if we focused on brane world models, our calculations can in principle be applied to any theory predicting a new physical mechanism which could be added in perturbation as a potential of the form $\kappa_b/r^b$. In particular, our upper bound (\ref{up}) is valid for any real number $b$.

\section{Conclusions}\label{conclu}
We studied the influence of power-law corrections to Newton's law on the gravitational quantum well spectrum. These corrections naturally arise in brane world models with infinite extra dimensions, and are generally less studied that the Yukawa ones. A particularity of the system we considered is that the interactions are not between pointlike particles, but between a pointlike particle and the Earth on one side, and between the particle and a mirror on the other side. We have thus computed the effective gravitational potential acting on the particle, where the brane world corrections are added in perturbation of the usual newtonian potential. Logically, the energy spectrum of our model, and consequently the critical heights, are affected by the corrections. In particular, the commonly predicted enhancement of the gravitational interaction at short distances causes the critical heights to be decreased: In a classical picture, the particle can bounce lower if the gravitation is stronger. 
\par Our results can be compared to those of the GRANIT experiment, which studies ultra cold neutrons in a gravitational quantum well. This comparison allows to obtain upper bounds on the typical length scale at which the correction becomes predominant in brane world models. These bounds are far from being as precise as those derived in the framework of LED models, but this could not be guessed before doing the calculations. This is due to the fact that the effective potentials are conversely proportional to the Earth's radius: The brane world contribution is thus very small in every case. We can conclude that an experiment such as GRANIT is nearly insensitive to such corrections, although its typical scale is the micrometer. Newton's law deformations should be better studied with pointlike bodies at the micrometer scale if one wants to get more precise bounds on the validity of the usual gravitational potential. 
\par Finally, let us point out that the existence of bound states in the GRANIT experiment is sufficient to make the following evident proposition: A theory forbidding bound states should be discarded. This leads us to suggest that a physical brane world model cannot predict that at long distance, the correction to Newton's law is more singular that $r^{-5}$. Indeed, the total effective potential acting on the neutron, i.e. $U_{b\geq5}(h)$ should be in this case more singular that $h^{-2}$. This criterion could serve as a test to discard unphysical approaches or to restrict the domain of validity of existing models.

\section*{Acknoledgments}
We are very much indebted to Prof. K. Protasov who initiated us into the GRANIT experiment, who gave us all the bases to pursue correctly in this way and with whom we had several fruitful discussions. F. B. thanks the FNRS Belgium, and V. M. thanks IISN for financial support.

\end{document}